\documentclass[12pt]{article}
\pdfoutput=1

\usepackage{amsmath,amsfonts,amssymb,amsthm,amssymb,bbm,bm}
\usepackage{pdfsync}
\usepackage{graphicx,import}
\usepackage[latin1]{inputenc}
\usepackage{hyperref}
\usepackage{empheq}
\usepackage{wrapfig}
\usepackage{float}
\usepackage[all]{xy}
\usepackage{stmaryrd}
\usepackage{rotating}
\usepackage{color}  
\usepackage{slashed}
\usepackage{cancel}
\usepackage{array, makecell} %
\usepackage{comment}
\usepackage{tikz}
\usepackage{hhline}
\usepackage{comment}
\usepackage{amsthm}
\usepackage{multicol}
\usepackage{cite}
\usepackage{setspace}
\usepackage{amsthm}
\usepackage{mathtools}
\usepackage{physics}
\usepackage{amsfonts}
\usepackage{xcolor}
\usepackage{graphicx}
\graphicspath{{plots/}}
\usepackage[left=23mm,right=13mm,top=35mm,columnsep=15pt]{geometry} 
\usepackage{adjustbox}
\usepackage{placeins}
\usepackage[T1]{fontenc}
\usepackage{lipsum}
\usepackage{csquotes}
\usepackage{hyperref}
\usepackage{tikz}
\usetikzlibrary{calc, arrows, decorations.pathmorphing, hobby}
\usepackage{subcaption}
\usepackage{graphicx}

\newcommand{\p}{\partial}

\newcommand{\cF}{{\mathcal F}}

\newcommand{\cO}{{\mathcal O}}

\newcommand{\be}{\begin{equation}}
\newcommand{\ee}{\end{equation}}
\newcommand{\bea}{\begin{eqnarray}}
\newcommand{\eea}{\end{eqnarray}}
\newcommand{\nn}{\nonumber}

\def\p{\partial}

\newcommand{\mean}[1]{\langle{#1}\rangle}

\def\a{\alpha}
\def\b{\beta}

\def\d{\delta}

\def\k{\kappa}
\def\m{\mu}
\def\n{\nu}

\def\vp{\varphi}

\def\om{\omega}

\def\Si{\Sigma}

\def\Om{\Omega}
\def\rd{\mathrm{d}}
\def\pa{\partial }

\newcommand{\scri}{\cal I}

\numberwithin{equation}{section}

\begin{document}
\unitlength = 1mm

\begin{center}
{\Large{\textsc{Hawking radiation far away from the event horizon}}}\\
\vspace{0.8cm}
Dawid Maskalaniec, Bart\l{}omiej Sikorski
\vspace{0.4cm}
\smallskip \\ \small{Faculty of Physics, University of Warsaw, Pasteura 5, 02-093 Warszawa, Poland}\\
\vspace{1cm}

\begin{abstract}
We study particle production in Vaidya spacetime. Using the WKB approximation, the distribution of Hawking radiation is calculated without the near-horizon approximation, which leads to finite corrections to the purely thermal spectrum. We extend our analysis to extremal and non-extremal Reissner-Nordstr\"om and Kerr black holes. Our results can be understood in terms of a thermodynamic toy model, where one regards Hawking radiation as Unruh radiation perceived by observers outside of the black hole. Moreover, we extend the model to incorporate the backreaction of Hawking quanta on spacetime geometry. Our study suggests that the backreaction may prevent the formation of the event horizon and spacetime singularity.
\end{abstract}
\end{center}

\thispagestyle{empty}

\newpage
\tableofcontents
\pagenumbering{arabic} 

\section{Introduction}
Fifty years ago Stephen Hawking discovered that black holes emit thermal radiation \cite{Hawking74,Hawking75}. Hawking's analysis of quantum fluctuations in a background of the gravitational collapse process, as already noted in \cite{Hawking75}, neglected trans-planckian modes \cite{unruh1995sonic,corley1996hawking,polchinski1995string}, backreaction of the radiated quanta on spacetime geometry \cite{susskind1992hawking,Kawai:2013mda}, and other possible quantum gravitational phenomena.

These problems motivated the search for alternative derivations of black hole radiance \cite{hartle1976path,davies1977radiation,robinson2005relationship,parikh2000hawking,shankaranarayanan2001method,chen2019hawking}. Although approaches like \cite{parikh2000hawking,shankaranarayanan2001method,chen2019hawking} significantly simplify the first Hawking's argument, and can be easily extended to a vast number of black hole spacetimes \cite{vanzo2011tunnelling}, it is not clear how they are related to the original picture \cite{helfer2019hawking}.

Thus, in the first part of this paper, we present yet another derivation of Hawking radiation, simplifying a little the original Hawking's calculation, but still working entirely within the realm of Quantum Field Theory in Curved Spacetime (QFTCS). We analyze scalar perturbations in Vaidya-like  (Schwarzschild, extremal and non-extremal Reissner-Nordstrom and Kerr-Vaidya) spacetimes. By gluing solutions of the Klein-Gordon equation across the shockwave of collapsing matter, we find Bogoliubov transformations between positive and negative frequency modes on past and future null infinities. For near-horizon Schwarzschild metric this approach was first
presented in \cite{ford1997quantumfieldtheorycurved}. We extend this idea to radiation further away from the horizon and to Reissner-Nordstrom-Vaidya and Kerr-Vaidya spacetimes. Conveniently, in Kerr spacetime, we do not have to use the argument of redefining energy and angular momentum in the ergosphere region to get the correct angular momentum dependence in the Boltzmann factor: $\exp[-\beta_H(\omega - m\Omega)]$.

 In Vaidya spacetime, the metric is constant in regions of positive/negative advanced time, so we interpret Hawking radiation in the Vaidya spacetime as created along the shockwave\footnote{In QFTCS the notion of a particle is nonlocal and can  be associated with a Cauchy surface. 
While particles we consider are defined on $\scri^\pm$, we emphasize that the shockwave is the only region where the dynamical nature of spacetime allows particle creation.} (i.e. "inside the collapsing star" for a more general collapse), rather than emitted from the black hole itself.  
In this context, Hawking radiation is a special case of the phenomenon of particle creation in a time-changing gravitational field, predicted by QFTCS \cite{Wald_book,birrell_davies_1982,parker2009quantum}. This brings to mind a remarkable possibility that black holes may not form \cite{unruh1976notes}. Hence, in the second part of the paper, we describe a simple, qualitative model of backreaction that leads to the so-called horizon-avoidance (recently studied in \cite{Kawai:2013mda,mersini2014backreaction,baccetti2018role,chen2018pre,unruh2018prehawking,barcelo2011minimal}).

\section{Hawking Radiation from a Vaidya Black Hole}
\subsection{Scalar field in Vaidya spacetime}
Let us first review the theory of a massless scalar field in a Vaidya spacetime:
\be\label{Vaidya}
\rd s^2 = -\left(1 - \frac{2M}{r}\Theta(v)\right)\rd v^2 + 2\rd v\rd r + r^2\rd\Om^2.
\ee
This metric describes a black hole formed by a shock wave (with stress tensor $T = -\frac{M}{4\pi r^2}\d(v)\rd v\otimes \rd v$). Free massless scalar field $\Phi$ in this spacetime satisfies the Klein-Gordon equation:
\be\label{KGeq}
\frac{1}{\sqrt{-g}}\p_{\m}(\sqrt{-g}g^{\m\n}\p_{\n}\Phi) = 0.
\ee
Upon separation of variables in regions $v\neq 0$:
\be
\Phi(x) = \frac{1}{r}\sum_{lm}\int_0^\infty \rd\om e^{-i\om t}R_{\om l}(r_{\star}) Y_{lm}(\theta,\vp),
\ee
where the tortoise coordinate $r_\star$ is defined via $\rd r_\star = \rd r/f(r)$ with $f(r) = 1-2M/r$ for $v> 0$ and $f=1$ for $v < 0$ and $t = v - r_{\star}$ is the standard Schwarzschild coordinate, the equation \eqref{KGeq} with metric \eqref{Vaidya} reduces to:
\be
\left(-\frac{\rd^2}{\rd r_{\star}^2} + V(r_\star)\right)R_{\om l} = \om^2 R_{\om l},
\ee
with effective potential $V(r_{\star}) = f(r)\left[\frac{l(l+1)}{r^2} + \frac{f'(r)}{r}\right]$. 

To simplify calculations, we shall consider only solutions with high frequency, i.e. we assume that $\omega^2 \gg V(r_{\star})$. This corresponds to a semi-classical solution with finite energy $E$ and frequency $\omega = E/\hbar$ (with $\hbar \rightarrow 0$). Then $R(r_{\star}) = e^{-i\omega r_{\star}} (1 + \cO(\om^{-1}))$.

We will associate states with two Cauchy surfaces for massless fields $\scri^-$ and $\scri^+\cup \mathcal{H}$ (with $\scri^\mp$ denoting past/future null infinity, and $\mathcal{H}$ black hole's horizon). On the past null infinity, $\scri^-$, modes
\be
p_{\om lm}(x) = \frac{1}{\sqrt{4\pi\om}}\frac{e^{-i\om t - i\om r_\star}}{r} Y_{lm}(\theta,\vp)
\ee
form a  basis of the space solutions to the Klein-Gordon equation, orthonormal with respect to the Klein-Gordon inner product:
\be
(p_{\omega lm}|p_{\omega' l'm'})=i\int_{\scri^-}\rd \Si^{\m}\;p_{\om'l'm'}^*\overset{\leftrightarrow}{\partial_\m} \,p_{\omega lm} = \delta_{ll'}\d_{mm'}\d(\omega-\omega'),
\ee
where $\rd \Si^\m$ is the volume element on $\scri^-$. Field decomposition w.r.t. the modes $p_{\omega lm}$ is:
\be
\Phi(x) = \int_0^{\infty}\rd\om \sum_{l = 0}^{\infty}
\sum_{m = -l}^{l} \left(a_{\om lm}p_{\om lm}(x) + a_{\om lm}^{\dagger}\bar{p}_{\om lm}(x)\right).
\ee
Operators $a_{\om lm}$ ($a_{\om lm}^{\dagger}$) annihilate (create) particles on $\scri^-$.

Since we are interested in the Hawking quanta emitted to infinity, we shall also single out modes outgoing to $\scri^+$:
\be
h_{\om lm}(x)= \Theta(r-2M)\frac{1}{\sqrt{4\pi\om}}\frac{e^{-i\om u}}{r} Y_{lm}(\theta,\varphi),
\ee
where $u=t-r_{\star}$ is the retarded Eddington-Finkelstein coordinate. These form an orthonormal basis of solutions of the KG equation that have the support on $\scri^+\cup \mathcal{H}$, limited to $\scri^+$. Corresponding field decomposition:
\be
    \Phi = \int_0^{\infty}\rd\om \sum_{l,m}\left(h_{\om lm}b_{\om lm}  + h_{\om lm}^{*}b_{\om lm}^{\dagger} \right) +\begin{pmatrix}
           \text{Part supported} \\
           \text{on the event horizon}
         \end{pmatrix}.
\ee
Operators $b_{\omega lm}$ ($b_{\omega lm}^\dagger$) annihilate (create) particles on $\scri^+$.

\subsection{Derivation of Hawking radiation}
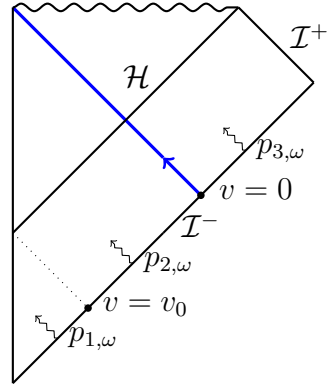
\begin{wrapfigure}{r}{6cm}
\begin{tikzpicture}[scale =1]
\coordinate (A) at (0,0); 
\coordinate (B) at (5,5); 
\coordinate (C) at (5,0); 
\coordinate (D) at (0,5); 
\coordinate (H) at (3,5); 
\coordinate (E) at (0,2); 
\coordinate (S) at (2,2); 
\coordinate (j) at (4,4); 
\draw[very thick, blue,->] (2.5,2.5)--(2,3);
\draw[very thick, blue] (2,3)--(0,5);
\draw[thick] (H) -- (j); 
\draw[thick] (A) -- (D);
\draw[thick] (A) -- (S); 
\draw[thick] (S) -- (j);
\draw[thick] (E) -- (H);
\draw[thick, decorate, decoration={snake, amplitude=0.5mm}] (0,5) -- (H);

\node at (2.6, 2.3) [above right] {$v=0$};
\node at (2.1,1.8)[above right] {$\scri^-$} ;
\node at (3.55,4.31)[above right] {$\scri^+$} ;
\node at (1.32,3.8)[above right] {$\mathcal{H}$} ;
\draw[dotted] (0,2) -- (1,1);
\node at (1.05, 0.75) [above right] {$v=v_0$};
\node at (1,1) [circle,fill,inner sep=1pt]{};
\node at (2.5,2.5) [circle,fill,inner sep=1pt]{};

\draw [->,line join=round,decorate, decoration={ zigzag, segment length=4,amplitude=.9,post=lineto, post length=2pt
}]  (0.6,0.6) -- (0.3,0.9);
\node at (0.6, 0.3) [above right] {$p_{1,\om}$};
\draw [->,line join=round,decorate, decoration={ zigzag, segment length=4,amplitude=.9,post=lineto, post length=2pt
}]  (1.6,1.6) -- (1.3,1.9);
\node at (1.6,1.3) [above right] {$p_{2,\om}$};
\draw [->,line join=round,decorate, decoration={ zigzag, segment length=4,amplitude=.9,post=lineto, post length=2pt
}]  (3.1,3.1) -- (2.8,3.4);
\node at (3.1,2.8) [above right] {$p_{3,\om}$};
 \end{tikzpicture}
\caption{Penrose diagram for Vaidya spacetime with marked wavepackets $p_{1\om}$, $p_{2\om}$, $p_{3\omega}$ localized on $\scri^-$ in regions $\{-\infty<v<v_0\}$, $\{v_0 <v < 0\}$ and $\{0<v<\infty\}$, respectively.}
\label{fig:penrose1}
\end{wrapfigure}
Hawking radiation comes from the relation of the state on $\scri ^-$ defined by operators $a_{\om lm}$ that represents the past without particles to the future state, outside the black hole, that is defined by operators $b_{\om lm}.$ 

The relation is encoded in Bogoliubov coefficients 
\begin{align}\label{bogotransf}
h_\om =\int_0^\infty \rd \om' \Big(\alpha_{\om\om'} p_{\omega'}+\beta_{\om\om'}p^*_{\omega'}\Big),
\end{align} 
where we dropped the $l,m$ indices as those will only contribute trivially to factors of $\delta_{ll'}\delta_{mm'}$.

To find $\a_{\om\om'}$, $\b_{\om\om'}$, let us split the incoming plane waves $p_{\om}$ as
\be
p_{\om} = p_{1,\om} + p_{2,\om} + p_{3,\om},
\ee
where
\begin{align}
\nn p_{1,\om} &= \Theta(-v-v_0)p_{\om},\\
\nn p_{2,\om} &= \Theta(v + v_0)\Theta(-v)p_{\om},\\
p_{3,\om} &= \Theta(v)p_\om,
\end{align}
with $v_0$ denoting the time after which all incoming waves will enter the event horizon (see Fig.\ref{fig:penrose1}). Only the part $p_{1,\om}$ will not fall into the black hole and it will pass $r=0$ to become an outgoing wave:
\be
p_{1\omega}\sim \frac{1}{r}e^{-i\om(t-r)} = \frac{1}{r}e^{-i\om (v-2r)}.
\ee

Note that, on the hypersurface $\{v=0\}$, modes $p_{i\omega}$ with $i=2,3$ are constant in $r$. Moreover, the continuity of solutions
\footnote{
The metric we consider is discontinuous, contrary to the usual form of junction conditions. This, however, does not cause discontinuity of solutions or scalar product on the surface $v=0$. This is because gluing takes place along the null surface and volume form is constant on null surfaces. The Klein-Gordon operator $\Box$, in coordinates $r,v,\theta,\varphi$, has discontinuous coefficients of $\partial_r^2,\partial_r$ but these are tangent to the surfaces of $v=$const.. In $\Box$, the derivatives of $v$ appear only in $2(\partial_r+1/r)\partial_v$, and $r$ coordinate is matched on both sides.
Therefore, any discontinuity of solutions in $v$, that would cause $c\delta(v)$ in values of solutions, cannot be compensated by discontinuity arising from the action of $\Box$.}
across the shockwave implies that $p_{1\omega}$ have the simple form from the flat space region: $p_{1\omega}(v=0) \sim \frac{1}{r} e^{-2i\omega r}$. Thus, pull-back of \eqref{bogotransf} on the shockwave is just a Fourier transform of $r\cdot h_{\omega}(v=0, r)$, with non-zero frequency Fourier modes $p_{1\omega}$ and zero-frequency part dependent on $p_{2\omega}, p_{3\omega}$:
\be
r\cdot h_{\omega} = \int_{-\infty}^{\infty}\rd\om' \left(\cF_{\om\om'}^{+}e^{2i\om' r} + \cF_{\om\om'}^{-}e^{-2i\om' r}\right).
\ee
Fourier coefficients with $\om'\neq 0$ are directly related to Bogoliubov coefficients:
\begin{align}
\alpha_{\om\om'} &= 4\pi\sqrt{\om'}\cF_{\om\om'}^{+}=\frac{2M}{\pi}\sqrt{\frac{\om'}{\om}}e^{4iM(\om - \om')}
    \left(-4iM(\om-\om')\right)^{-1-4iM\om} \Gamma \left(1 +4iM\om\right),
\\
\beta_{\om\om'} &= 4\pi\sqrt{\om'}\cF_{\om\om'}^{-}=\frac{2M}{\pi}\sqrt{\frac{\om'}{\om}}e^{4iM(\om + \om')}
    \left(-4iM(\om+\om')\right)^{-1-4iM\om} \Gamma \left(1 +4iM\om\right).
\end{align}
We can make sense of the factor $(-i)^{-2ir_s\om}$ with a regularization $\lim_{\epsilon\rightarrow0 }(\om'\mp i\epsilon)r$ in the place of $\om'r$. 

Note that the above coefficients do not satisfy the relation $|\alpha_{\om\om'}| = e^{\pi\omega/\kappa}|\beta_{\om\om'}|$ (with $\kappa$ being the surface gravity of the event horizon), which is the hallmark of the thermal spectrum of Hawking quanta \cite{Hawking75}. Hence, we need to directly calculate the number of quanta on $\scri^+$ in the vacuum on $\scri^-$, $|0_{\scri^-}\rangle$:
\be\label{N_from_beta_divergent}
\mean{N_{\om}^{\scri^{+}}}= \langle 0_{\scri^-}|b_{\om}^{\dagger}b_{\om}|0_{\scri^-}\rangle =\int_{0}^{\infty}|\b_{\om\om'}|^2\rd\om'
= \frac{e^{-2\pi r_s \om}}{\pi^2\om}  \left|\Gamma \left(1 +2ir_s\om\right)\right|^2\int_{0}^{\infty}\frac{\om'\rd\om'}{4(\om+\om')^2}.
\ee
The integral is logarithmically divergent and leads to an infinite value of $\mean{N_{\om}^{\scri^{+}}}$. The reason for this, due to the Shale theorem \cite{Shale1962LINEARSO}, is that vacua on $\scri^+$ and on $\scri^-$ are not unitarily equivalent. To make sense of the result, one needs to introduce a regularization. A finite result can be obtained if we use the UV cutoff $\Lambda$ during the integration in $\om'$
\begin{align}\label{NSchw}
    \mean{N_{\om}^{\scri^{+}}}=\frac{2 M}{\pi}\frac{1}{e^{8\pi M\om}-1}\left(\log\left(\frac{\Lambda}{\om}\right) - 1 + \cO(\Lambda^{-1})\right),
\end{align}
where the leading part in $\Lambda$ is the expected black body spectrum of the Hawking temperature $T_H=\frac{1}{8\pi M}$. Since we have not taken the near-horizon limit, we have a correction $\sim\log(\om)$ due to Hawking quanta created at early times, before the horizon has formed. If the cutoff $\Lambda$ does not depend on $\omega$, this term cannot be cancelled by a cutoff redefinition. 

The final result depends on the regularization but, importantly, the leading term is the number of particles in thermal distribution which is not finite but corresponds to a well-defined state. As we shall see in the section \ref{sec:thermo_int}, the logarithmic divergence is equivalent to the use of the position cutoff resembling the regularization method introduced by 't Hooft \cite{hooft1985quantum}.
 The $\log\om$ term amounts to the correction to the total number of particles that is UV finite.
Therefore, the infinite contribution to Hawking radiation is thermal, just as was argued in \cite{Hawking75}, and it agrees with the general result of Fredenhagen and Haag \cite{FredenhagenHaag}.

The method of calculation of the Bogoliubov coefficients by gluing solutions across the shockwave can be easily generalized to any spherically symmetric Vaidya-like black hole background:
\begin{align}\label{sphericalSym_f(r)_metric}
\rd s^2 = -f(r, v)\rd v^2 + 2\rd r^2\rd r + r^2 \rd\Om^2,
\end{align}
with $f(r, v>0) = f(r)$ and $f(r, v<0) = 1$. For example, for non-extremal Reissner-Nordstr\"om black hole we obtain (see \hyperref[Appendix_RN]{Appendix} for the derivation):
\begin{align}
\nn\beta_{\om\om'} =&\frac{1}{\pi}\sqrt{\frac{\om'}{\om}} \Big(\frac{r_+^2}{ r_+-r_-}\Big)^{ -\frac{i\om}{2\kappa_+}}\Big(\frac{r_-^2}{ r_+-r_-}\Big)^{ -\frac{i\om}{2\kappa_-}}e^{i(\om+\om')(r_+ + r_-)} \big(-2i(\om+\om')\big)^{-1-\frac{i\om}{2\kappa_+}-\frac{i\om}{2\kappa_-}} \\
&\times \Gamma\Big(1+\frac{i\om}{\kappa_+}\Big) W_{\frac{i\om}{2}\left(\frac{1}{\kappa_-} - \frac{1}{\kappa_+}\right), -\frac{1}{2}- \frac{i\om}{2}\left(\frac{1}{\kappa_+} + \frac{1}{\kappa_-}\right)}\Big(-2i(r_+ - r_-)(\om + \om')\Big),
\end{align}
where $r_{+}, \kappa_+$ ($r_-$, $\kappa_-$) are the coordinate radius and surface gravity of the outer (inner) horizon. $W_{\alpha,\beta}(z)$ is the Whittaker function \cite{ryzhikSpecial}. The expectation value of the particle number operator is:
\be\label{ReissnerNumber}
\mean{N_{\om}^{\scri^{+}}} = \frac{1}{2\pi\kappa_+}\frac{\log (\Lambda/\omega)}{e^{2\pi\om/\kappa_+}-1}  + \cO(\Lambda^0).
\ee

\subsection{Near-horizon approximation}
If one takes the near-horizon limit before calculating the Bogoliubov coefficients, one replaces $h_{\om}$ with:
\be
h_{\om}^{(NH)}=\frac{\Theta(r- 2M)}{4\pi\sqrt{\om}}e^{-i\om v}\exp\left[4iM\om + 4iM\om \log(r/2M-1)\right].
\ee
This is of course a valid approximation for wavepackets localized near the horizon, where $2i\omega r$ is slowly varying compared to the logarithm. Now, if we repeat the calculation of Bogoliubov coefficients for $h_{\om}^{(NH)}$, we obtain standard expressions \cite{Hawking75}:
\begin{align}\label{NHbogoliubov}
    \alpha_{\om\om'}^{(NH)} &=\frac{2M}{\pi}\sqrt{\frac{\om'}{\om}}e^{4iM(\om - \om')}
    \left(4iM\om'\right)^{-1-4iM\om} \Gamma \left(1 +4iM\om\right),
    \\
    \beta_{\om\om'}^{(NH)} &= \frac{2M}{\pi}\sqrt{\frac{\om'}{\om}}e^{4iM(\om + \om')}
    \left(-4iM\om'\right)^{-1-4iM\om} \Gamma \left(1 +4iM\om\right).
\end{align}
Choosing, as before, $(-1)^{-4iM\om} = e^{-4\pi M\om}$ we get:
\be 
|\alpha_{\om\om'}^{(NH)}|=e^{2\pi r_s\om}| \beta_{\om\om'}^{(NH)}|,
\ee
which yields the thermal distribution.

For a general spherically-symmetric Vaidya black hole, assuming non-extremality, at the horizon $r=r_s$, $f(r=r_s) = 0$ we have $f'(r=r_s)=2\kappa\neq 0$, with $\kappa$ denoting the surface gravity. Then, we can repeat the calculation with $h_\om \sim e^{2i\om r_\star}$, $p_\om \sim e^{2i\om r}$ at the shockwave, and tortoise coordinate $r_\star$ defined via $\rd r_{\star} = \rd r/f(r)$. In the near-horizon limit, we consider only leading terms in $r-r_s$ so that 
\be
f(r)=2\kappa (r-r_s)+\cO((r-r_s)^2), \qquad r_\star=\frac{1}{2\kappa}\log\left(\frac{r}{r_s}-1\right) +\cO(r-r_s).
\ee
Near the horizon modes $h_{\omega}$ can be approximated by
\be
h_{\om}^{(NH)} = \frac{\Theta(r-r_s)}{\sqrt{4\pi\om}}e^{-i\om v}e^{i\frac{\om}{2\kappa}\log\left(\frac{r}{r_s} - 1\right)}.
\ee
Resulting Bogoliubov coefficients are given by the same expressions as \eqref{NHbogoliubov}, but with $4M$ suitably replaced by either $2r_s$ or $1/\kappa$, so that they satisfy $|\alpha_{\om\om'}^{(NH)}|=e^{\frac{2\pi\om}{\kappa}}| \beta_{\om\om'}^{(NH)}|$, from which we recover the thermal distribution with the Hawking temperature $T_H=\frac{\kappa}{2\pi}$.

\section{Extremal Reissner-Nordstr\"om black hole}
From \eqref{ReissnerNumber} we can see that in the extremal limit of the Reissner-Nordstr\"om (RN) black hole the divergent part of the particle number vanishes for every $\om>0$, since then $\kappa_+\rightarrow 0$. 
There is no infinite particle production near the extremal horizon.
Nevertheless, one may expect that a finite number of Hawking quanta can be produced further away from the horizon. 
Thus, let us analyze the particle production in the extremal Vaidya-RN spacetime:
\be
\rd s^2 = -\left(1 - \frac{r_s}{r}\Theta(v)\right)^2 \rd v^2 + 2\rd r\rd v + r^2\rd\Om^2. 
\ee
Even though the formation of an extremal black hole may seem to be an unphysical process (since strong electric repulsion should stop the collapse process), it was recently shown that extremal RN black holes can form from critical Einstein-Vlasov systems \cite{kehle2024extremal}. 

As before, we decompose the outgoing high-frequency modes in the black hole region ($v > 0$), localized outside of the horizon $r = r_s$:
\begin{align}
h_{\om} = \frac{\Theta(r-r_s)}{4\pi\sqrt{\om}}\frac{1}{r}e^{-i\om v + 2i\om r_{\star}},
\end{align}
where the tortoise coordinate is
\begin{align} r_{\star} = \int \frac{\rd r}{(1-r_s/r)^{-2}} = r - r_s^2/(r-r_s) + 2r_s\log(r/r_s - 1),\end{align}
into the sum of outgoing modes in the Minkowski region:
\be
p_{1\om} = \frac{\Theta(r-r_s - v/2)}{4\pi\sqrt{\om}}\frac{1}{r}e^{-i\om v + 2i\om r}.
\ee
This gives us Bogoliubov coefficients:
\be
    \beta_{\om\om'} =\frac{2}{\pi}\sqrt{\frac{\om'}{\om}} r_s^{-4i\om r_s} e^{-2i(\om+\om')r_s}\left(\frac{-\om }{\om + \om' }\right)^{2ir_s \om + 1/2} K_{1 + 4i r_s \om}\left(4r_s\sqrt{\om(\om + \om')}\right),
\ee
where $K_{\a}(z)$ is the modified Bessel function of the second kind. The expectation value of the number operator,
\begin{align}\label{NuvERNBH}
\mean{N_{\om}^{\scri^+}} = \frac{4r_s^2}{\pi^2}\int_{0}^{\infty}\rd\om' \frac{\om'}{\om + \om'} \left|K_{1 + 4ir_s \om}(4r_s\sqrt{\om(\om + \om')})\right|^2,
\end{align}
is finite since the integrand is bounded for $\om'\rightarrow0$ and vanishes exponentially at large $\om$ because $K_\alpha(z)\sim\sqrt{\frac{\pi}{2z}}e^{-z}$ at large z.

As argued in \cite{Hawking75}, divergence in $\mean{N_{\om}^{\scri^+}}$ implies that a black hole must evaporate due to a constant energy flux at late times at $\scri^+$. Here, a finite $\mean{N_{\om}^{\scri^+}}$ means that the energy flux of Hawking radiation at $\scri^+$ is dispersed at late times. There is no thermal radiation flux and hence, no reason for the extremal black hole to evaporate.

These results agree with the analysis of particle production between two accelerating mirrors corresponding to an extremal RN black hole \cite{Angheben:2005,good2020extremal}.

\section{Kerr Black Hole Radiance}
In this section, we shall extend our analysis to Kerr-Vaidya spacetime. To fix the notation, let us first recall basic facts about the Kerr solution.

Kerr metric in advanced Eddington-Finkelstein coordinates $(v,r,\theta,\phi)$ reads:
\be
    \mathrm{d}s^2 =-\frac{\Delta}{\Sigma}(\mathrm{d}v-a\sin^2\theta\mathrm{d}\phi)^2 + 2\rd r(\rd v - a \sin^2\theta \rd\phi)
    + \Sigma \mathrm{d}\theta^2 + \frac{\sin^2\theta}{\Sigma}(\rho\mathrm{d}\phi - a \mathrm{d}v)^2,
\ee
where $\Delta = r^2 - 2Mr + a^2,$ $\rho = r^2 + a^2$, $\Sigma = r^2 + a^2 \cos^2\theta$ and $a =J/M$ is the angular momentum per unit mass. Outer and inner horizons are located at $r_{\pm} = M\pm \sqrt{M^2} \pm \sqrt{M^2 - a^2}$, $\Delta(r) = (r-r_+)(r-r_-)$. Boyer-lindquist coordiates $(t,r,\theta,\vp)$ are defined via: $\rd t = \rd v - \rho \rd r/\Delta$ and $\rd \vp = \rd\phi - a\rd r/\Delta$.

The simplest non-stationary generalization of the Kerr metric is obtained by introducing a time-dependent mass, $M = M(v)$. We shall consider $M(v) = M\cdot\Theta(v)$, where $M=\text{const.}$ and $\Theta(v)$ is the Heaviside step function. Such a metric can be obtained from the Newman-Janis trick \cite{NewmanJanis1965} with Vaidya metric \eqref{Vaidya} as the seed metric \cite{Dahal:2020KerrVaidya}. This metric is over-extreme for $v < 0$, but, as we will argue later, our method is valid also for a wider class of Kerr-Vaidya collapses. However, it is generically hard to model sources of axially symmetric gravitational field. Thin disk solutions \cite{Bicak:1993ThinDisk} are among a few known more realistic solutions with exterior Kerr metric.

Boyer-Lindquist coordinates in the region $v > 0$, $(t^>, r,\theta, \vp^>)$ are:
\be
t^> = v - \frac{1}{2\kappa_+}\log\left(\frac{r}{r_+} - 1\right) - \frac{1}{2\kappa_-}\log\left(\frac{r}{r_-} - 1\right),\quad \vp^> = \phi - \frac{a}{r_+ - r_-}\log\left(\frac{r-r_+}{r-r_-}\right).
\ee
For $v < 0$ we accordingly have:
\be
t = v - r,\quad \vp^< = \phi - \arctan\left(\frac{r}{a}\right).
\ee

Let us focus on the outer region, $v>0$, where we have the non-extremal Kerr metric. The Klein-Gordon equation in Kerr spacetime reduces, after a separation of variables:
\be
    \Phi(x) = e^{-i\omega t}e^{im\phi} S_{\omega lm}(\cos(\theta))R_{lm\om}(r),
\ee
to \cite{carter1979generalized,teukolsky1973perturbations,press1973perturbations,teukolsky1974perturbations}:
\begin{align}\label{KGeqKerr}
    \nn\Big[\frac{\mathrm{d}}{\mathrm{d}r}\Delta \frac{\mathrm{d}}{\mathrm{d}r} + \frac{(2M r_+\omega - am)^2}{(r - r_+)(r_+ - r_-)} & - \frac{(2M r_-\omega - am)^2}{(r - r_-)(r_+ - r_-)} 
    \\
    &+ (r^2 + 2M(r+2M))\omega^2\Big]R(r) = K_{\omega lm}R(r),
\end{align}
where $r_\pm = M\pm\sqrt{M^2 - a^2}$ and $S_{\omega lm}(x)$ are the spheroidal harmonics \cite{AbramowitzStegun}. In general, separation constants $K_{\omega lm}$ are known only numerically. In WKB approximation, $\omega \gg M^{-1}, a$, with large angular momentum $m\gg 1$, solutions to Eq.~\eqref{KGeqKerr} read:
\be\label{WKBSolKerr}
    R_{\pm}^{>}(r) = \exp\Bigg[\pm i\int \frac{\mathrm{d}r}{\Delta(r)}\sqrt{(2Mr_+\om - am)^2 \frac{r-r_-}{r_+-r_-} + F(r)}\Bigg],
\ee
where by $F(r)$ we denoted a contribution which vanishes at the outer horizon:
\be
F(r) = \Delta(r)\left[ (r^2 + 2M(r + 2M))\om^2 -K_{\om lm} \right]- (2Mr_-\om - am)^2 \frac{r-r_+}{r_+ - r_-} .
\ee
Since separation constants $K_{\omega lm}$ do not depend on the mass $M$, solutions in the region $v < 0$, $R_{\pm}^{<}$, are given by the same expression but with $\Delta(r)$ replaced by $r^2 + a^2$ and $M=0$:
\be
R^<_{\pm}(r) = \exp\left[\pm i\int\frac{\rd r}{r^2 + a^2}\sqrt{a^2 m^2 + (r^2 + a^2)(r^2 \om^2 - K_{\omega lm})}\right].
\ee
Hence, the outgoing part of the mode $p_{\om}$, $p_{1\om}$, (normalized w.r.t. the Klein-Gordon inner product to the Dirac delta) in the region $v < 0$ is:
\be\label{p1omKerr}
p_{1\om} = \frac{\Theta(r-r_+)}{\sqrt{4\pi\om}}\frac{1}{r}S_{lm\om}(\cos\theta)R_+^{<}(r)e^{-i\om v + im\phi}e^{i\om r - im\arctan(r/a)}.
\ee
Normalized modes outgoing to $\scri^+$ are:
\be\label{homKerr}
h_{\om} = \frac{\Theta(r-r_+)}{\sqrt{4\pi\om}}\frac{1}{r}S_{lm\om}(\cos\theta)R_+^{>}(r)e^{-i\om v + im\phi}\exp\left[i\om r - \frac{ima}{r_+ - r_-}\log\left(\frac{r-r_+}{r-r_-}\right)\right].
\ee
Bogoliubov transformation between $h_{\om}$ and $p_{\om}$ can be determined by calculating the Klein-Gordon inner product of $p_{\om}$ and $h_{\om}$ on the Cauchy slice consisting of the shockwave and the part of past null infinity from the shockwave to the spatial infinity $i^0$:
\be
\beta_{\om \om'} = -i\int_{\Si_1}\rd \Si^\m p_{\om'}\overset{\leftrightarrow}{\pa_\m}h_{\om} - i\int_{\Si_2}\rd \Si^\m p_{\om'}\overset{\leftrightarrow}{\pa_\m}h_{\om},
\ee
where $\Si_{1} = \{(v,r,\theta,\phi)|v=0\}$ and $\Si_{2}=\{(v,r,\theta,\phi)|r\rightarrow \infty,\; v > 0\}\in \scri^-$. Since $h_{\om}\sim e^{-i\om u}$ is constant at $\scri^-$, the integral over $\scri^-$ gives zero for $\om'\neq 0$. Since $h_{\om}(r < r_+)=0$, we have:
\be
\beta_{\om\om'} = -i\int_{r_+}^{\infty}\rd r \int \rd\Om\, p_{\om'}\overset{\leftrightarrow}{\pa_\m}h_{\om},
\ee
where $\rd\Om$ is the volume form of the round unit 2-sphere. In the part of the shockwave outside of the event horizon $p_{2\om}$, $p_{3\om}$ are constant, so only $p_{1\om}$ gives a non-zero contribution:
\be\label{betaKerr}
\beta_{\om\om'} = -i\int_{r_+}^{\infty}\rd r\int \rd\Om\, p_{1\om'}\overset{\leftrightarrow}{\pa_\m}h_{\om},
\ee
with $p_{1\om'}$, $h_{\om}$ given by \eqref{p1omKerr} and  \eqref{homKerr} evaluated at $v=0$.

The explicit formula is not very illuminating but it is worth noting that it gives the correct near-horizon limit. For $r\rightarrow r_+$ we have:
\begin{align}
h_{\om}(x)\sim \frac{\Theta(r-r_+)}{\sqrt{4\pi\om}}&\frac{S_{lm\om}(\cos\theta)}{r_+}\left(\frac{r}{r_+}-1\right)^{i(\om - m\Om_+)/\kappa_+}e^{-i\om v+im\phi + i\om r_+},
\\
\nn p_{1\om}(x)\sim \frac{\Theta(r-r_+)}{\sqrt{4\pi\om}}&\frac{S_{lm\om}(\cos\theta)}{r_+}e^{-i\om v + im\phi}\exp\Bigg[i\om r - im\arctan\left(\frac{r}{a}\right)+ 
\\
&+ \frac{i(r-r_+)}{r_+^2 + a^2}\sqrt{a^2 m^2 + (r_+^2 + a^2 )(r_+^2 \om^2 - K_{\om lm})}\Bigg],
\end{align}
where $\Om_+ = \frac{a}{2Mr_+}$ is the angular velocity of the outer horizon. To calculate the divergent part of the expectation value of the particle number, we need to take only terms that dominate at large $\om'$. From \eqref{betaKerr} we obtain:
\be
\beta_{\om\om'}^{(NH)}\overset{\om'\rightarrow \infty}{\sim} \frac{-i r_+}{4\pi\sqrt{\om \om'}} \left[-i\om'\left(1 + \frac{r_+}{\sqrt{r_+^2 + a^2}}\right)\right]^{- i(\om - m\Omega_+)/\kappa_+}\Gamma\left(1 + \frac{i(\om - m\Omega_+)}{\kappa_+}\right),
\ee
so the particle number
 \begin{align}\label{kerrnumber}
    \mean{N_{\om m}^{\scri^+}}= \frac{r_+^2}{16\pi \kappa_+}\frac{\om -m\Om_+}{\om} \frac{1}{e^{2\pi(\om - m\Om_+)/\kappa_+} - 1}\left(\log \Lambda+ \cO(\Lambda^0)\right).
\end{align} 
Note that in the near-horizon limit, the WKB approximation is valid for superradiant modes, i.e. for $\om - m\Om_+ < 0$, so the above result also holds in this case. Result \eqref{kerrnumber} is in total agreement with the analysis of superradiant modes of \cite{MaldacenaStrominger1997}.

\section{Thermodynamic interpretations}\label{sec:thermo_int}
In \cite{Jacobson:1995ak}, Jacobson noticed that the Schwarzschild black hole Hawking temperature coincides with the redshifted Unruh temperature (measured by a distant observer) of a box at fixed position $(r,\theta,\vp)$ close to the event horizon. Acceleration of the box at position $r$ is:
\be\label{boxacceleration}
a(r) = \frac{M}{r^2}\left(1-\frac{2M}{r}\right)^{-1/2}.
\ee
\begin{wrapfigure}{r}{6cm}
\begin{tikzpicture}[scale =1, use Hobby shortcut]
\coordinate (A) at (0,0); 
\coordinate (B) at (5,5); 
\coordinate (C) at (5,0); 
\coordinate (D) at (0,5); 
\coordinate (H) at (2,5); 
\coordinate (E) at (0,3); 
\coordinate (S) at (2,2); 
\coordinate (j) at (3.5,3.5); 
\draw[very thick, blue,->] (2.5,2.5)--(2.2,2.8);
\draw[very thick, blue] (1.9,3.1)--(2.2,2.8);
\draw[very thick, blue] (1.5,3.5)--(0,5);
\draw[very thick, red] (1.5,3.5)--(1.9,3.1);

\draw[thick] (H) -- (j); 
\draw[thick] (A) -- (D);
\draw[thick] (A) -- (S); 
\draw[thick] (S) -- (j);
\draw[thick] (E) -- (H);
\draw[thick, decorate, decoration={snake, amplitude=0.5mm}] (0,5) -- (H);

\node at (2.1,1.8)[above right] {$\scri^-$} ;
\node at (2.85,4.01)[above right] {$\scri^+$} ;
\node at (1.25,3.8)[above right] {$\mathcal{H}$} ;

\draw[red,thick] (2,3.2)-- (1.8,3);
\draw[red,thick] (1.4,3.4)-- (1.6,3.6);
\draw[red,thick,->] (1.45,3.25)-- (1.35,3.35);
\draw[red,thick,->] (1.45,3.25)-- (1.75,2.95);
\node at  (1.55,3.15)[red, above right] {\small $T(r_i)$} ;
\node at  (1.58,3.35)[red, below left] {\small $\delta r$} ;
\end{tikzpicture}
\caption{Penrose diagram of Vaidya spacetime. In red we have marked a box at position $r_i$, of width $\delta r$ (in the affine parameter $r$). It contains Unruh radiation of temperature $T(r_i) = a(r_i)/2\pi$ with $a(r_i)$ given by \eqref{boxacceleration}.}
\label{fig:UnruhBoxes}
\end{wrapfigure}
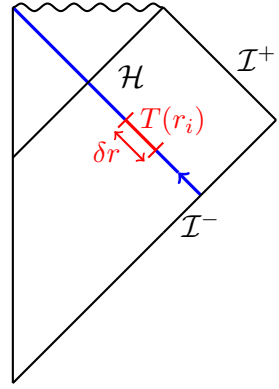

The corresponding Unruh temperature reads $T_U(r) = a(r)/2\pi.$ To a distant observer (sitting at $r\rightarrow\infty$), the temperature of the box would appear redshifted by a factor $(1-2M/r)^{1/2}$:
\be
T_{obs}(r) = \frac{M}{2\pi r^2},
\ee
so for $r = 2M$, $T_{obs}$ matches the Hawking's result, $T_{obs}(r=2M) = 1/8\pi M$.

In this and the subsequent section, we investigate this idea and consider Hawking radiation as Unruh radiation in a box travelling through the Vaidya spacetime. 
In this case, excited plane wave mode, $h_{\om}$, should be just a collection of $N\gg 1$ boxes filled with photon gas, evenly distributed along the shockwave $v=0$. Set the $i$'th box to be localized in the region $(r_i - \d r/2, r_i + \d r/2)$ (see Fig. \ref{fig:UnruhBoxes}). The free energy of the $i$'th box is

\be\label{deltaF}
\d F_i = -T(r_i)\log(Z_i) = -T(r_i)\int_0^{\Lambda}\rd\om \;\d\rho_i(\om)\log(1 - e^{-\om/T(r_i)}),
\ee
where $\d\rho_i(\om)\rd\om$ is the density of states. Generically, it will depend on some power $p$ of $\om$ and be extensive in the size of the box, $\d r$. Thus:
\be
\d\rho_i(\om)\rd\om = C\om^p\,\d r\,\rd \om,
\ee
with some constant $C$. We need some length measure to make $\d\rho_i(\om)\rd\om$ dimensionless. Summing over all boxes gives us the total free energy of the Hawking radiation:
\be
F = \sum_i \d F_i = -C \int_{2M}^{\infty}\rd r \,T_U(r)\int_0^\Lambda\rd\om\;\om^p \log\left(1-e^{-\om /T_U(r)}\right).
\ee
For now, let us choose $p=0$. Then the above integral is logarithmically divergent in $\Lambda$. Upon changing the integration variable $\om \mapsto \om T(r)/T_H$ (where $T_H$ is the Hawking temperature) and changing the order of integration:
\be
\int_{r_s}^{\infty}\rd r\int_0^{\Lambda}\rd\om = \int_{0}^{\infty}\rd\om\int_{r_s + b(\om)}^{\infty} \rd r\cdot \frac{T(r)}{T_H},
\ee
we obtain
\be
F = -\frac{C}{T_H}\int_0^{\infty}\rd\om\int_{r_s+b(\om)}^{\infty}\rd r\;T(r)^2\log(1 - e^{- \om/T_H}).
\ee
UV-cutoff in position space, $b(\om)$, is determined by $\Lambda = \om T_H/(T(r_s + b(\om)))$. In the leading order:
\be\label{spacecutoff}
b = r_s \cdot \left(\frac{\om}{\Lambda}\right)^2.
\ee
Thus,
\be
F = -\text{const.}\times\int_0^\infty\rd\om\;\log\left(\frac{\Lambda}{\om}\right)\log\left(1 - e^{-\om/T_H}\right) + \cO(\Lambda^0).
\ee
We arrived at an effective thermodynamic system at temperature $T_H = 1/4\pi r_s$, with density of states:
\be\label{eff_dos}
\rho(\om)\rd\om \propto \log\left(\frac{\Lambda}{\om}\right)\rd\om.
\ee
With this density of states, the expectation value of the particle number:
\be\label{NUnruhBox}
\mean{N} \sim \int_0^{\infty}\rd\om\,\log\left(\frac{\Lambda}{\om}\right)\left(e^{\om/T_H} - 1\right)^{-1}
\ee
agrees with the result \eqref{NSchw}. This resembles the "brick-wall model" introduced by 't~Hooft \cite{hooft1985quantum}. The divergent part of the radiation comes from the region comes from the region near stretched horizon $r=2M +b(\om)$. If $\Lambda^{-1}$ is of the order of the Compton wavelength of the black hole, then $b$ represents uncertainty in the position of the horizon.

Note that the integral \eqref{deltaF} is convergent for $\Lambda \rightarrow \infty$, so we could alternatively introduce fixed position space UV cutoff $b$ while integrating over $r$. With $b$ independent of $\om$, $\Lambda/\om = \sqrt{2M/b}$, and the spectrum \eqref{NUnruhBox} becomes purely thermal in the leading order. 

One can check that the divergent part of the result is independent of trajectories of the boxes as long as they do not fall into the black hole. A worldline infinitesimally close to a non-extremal event horizon (at $r=r_H$) must have an infinite acceleration $\alpha\sim (r-r_H)^{-1/2}$, and thus, infinite Unruh temperature. Thus, for appropriate density of states, free energy $F\sim \int_{r_H + b}^{\infty}\rd r T(r)^{2}$ will be logarithmically divergent. 

Interestingly, worldlines nearby an extremal horizon do not have infinite accelerations. E.g., for extremal RN black hole worldlines of constant $r,\theta,\vp$ have acceleration equal to $1/M$ for every value of $r>r_H$, so that corresponding free energy is divergent in infrared rather than UV.

\section{A simple backreaction model}
It the Vaidya spacetime positive and negative-frequency modes (w.r.t. the advanced time $v$) mix only on the shockwave hypersurface. Hence, if we extend the notion of particles from the asymptotic region into the bulk using this notion of positive-frequency modes, 
 it is possible to treat
 Hawking quanta as produced on the shockwave hypersurface, before the event horizon forms. 
This might imply that black holes may not even form - they could "evaporate before they are formed". 

This idea, firstly put forward already in the seminal work of Unruh \cite{unruh1976notes}, became an interest of a number of recent papers \cite{Kawai:2013mda,mersini2014backreaction,baccetti2018role,chen2018pre,unruh2018prehawking,barcelo2011minimal,Barcelo:2010PreHawkingMinimalConditions}. In \cite{Barcelo:2010PreHawkingMinimalConditions} Barcelo~\textit{et.al.} argued that a large thermal flux of so-called "pre-Hawking quanta" can be produced near a horizonless object. However, findings of Unruh~\textit{et.al.}, based on analysis of stress-energy tensor in simplified models of the black hole formation, suggest that pre-Hawking radiation might not be enough for a star to evaporate before the event horizon forms \cite{chen2018pre,unruh2018prehawking}. Nevertheless, since other models were also proposed \cite{mersini2014backreaction,mersini2014back,Kawai:2013mda}, the status of pre-Hawking radiation is unclear. 

To illustrate horizon avoidance, let us extend the thermodynamic model from the previous section. A simple way to model backreaction is to assume that the shockwave has to lose energy when Hawking quanta are emitted (boxes with Unruh radiation are created). When the shockwave travels affine distance $\delta r$, its energy decreases by the energy of one box with Unruh radiation (with a density of states proportional to $\om^p$):
\be\label{backreaction}
\delta M = -\d r\cdot C\int_0^{\infty}\rd\om\,\frac{\om^{p+1}}{e^{\om/T_U(r)} - 1}.
\ee
We have taken an infinite UV cutoff, $\Lambda\rightarrow \infty$, expecting that the horizon will not form, so there will be no UV divergences. In this way, the Bondi mass aspect $M$ becomes $r$-dependent. We assume that the metric is of the form:
\be
\rd s^2 = -\left(1- \frac{2M(r)}{r}\Theta(v)\right)\rd v^2 + 2\rd r\rd v + r^2 \rd\Omega^2. 
\ee
Eq. \eqref{backreaction} implies self-consistency equation for $M(r)$:
\be\label{backreaction2}
M(r) = M_{\infty} + C\int_{\infty}^{r}\rd r'\int_0^{\infty}\rd\om \frac{\om^{p+1}}{e^{\om/T(r')}-1} = M_{\infty} + C\,\text{Li}_{p+2}(1)\Gamma(p+2) \int_{\infty}^{r}\rd r\,(T(r))^{p+2},
\ee
where $M_{\infty} = M(r=\infty)$, $\text{Li}_{s}(z)$ is the polylogarithm function, and
\be
T(r) = \frac{1}{4\pi}\frac{f'(r)}{\sqrt{f(r)}},\quad f(r) = 1 - \frac{2M(r)}{r}.
\ee

\begin{figure}
\centering
    \includegraphics[scale = 0.7]{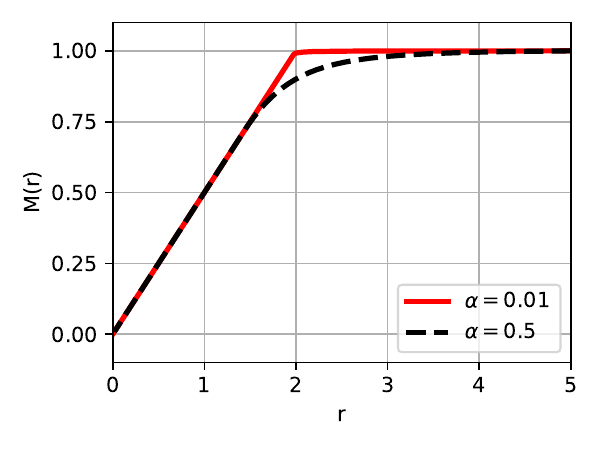}
    \caption{Plots of solutions to equation \eqref{br3} for initial energy of the shockwave $M_\infty = 1$ and backreaction parameters $\a = 0.5$ (dashed black line)  and $\a = 0.01$ (solid red line).}
    \label{fig:plot}
\end{figure}

We can fix the values of $p$ and $C$ e.g. by requiring that the entropy of all boxes with radiation,
\be
S = -\sum_i \frac{\pa\d F_i}{\pa T(r_i)}
\ee
matches the Bekenstein-Hawking entropy, $S = 4\pi M_{\infty}^2$. However, such a condition cannot be solved analytically. Thus, for simplicity, we shall assume $p=0$, as in the previous section. Then, differentiating \eqref{backreaction} once, we obtain:
\be
r^2 (r-2M(r))M'(r) = \alpha (M(r) - rM'(r))^2,
\ee
with $\alpha = C\,\text{Li}_2(1)/16\pi^2$. At $\alpha = 0$ we recover the classical limit, $M'(r)=0$. One branch of solutions:
\be\label{br3}
M'(r) = \frac{M}{r} - \frac{M}{\alpha} + \frac{r}{2\a} - \sqrt{\left(\frac{M}{\alpha} -\frac{r}{2\alpha} -\frac{M}{r} \right)^2 - \frac{M^2}{r^2}},
\ee
gives us the expected behavior of Bondi mass $M(r)$. Figure~\ref{fig:plot} shows numerical solutions to this equation for $M_{\infty}=1$ and $\alpha = 0.5$ and $\alpha = 0.01$. In the first case, backreaction effects are strong and we can see that $M(r)$ starts to decrease significantly before $r = 2$ and the event horizon does not form. At $r=0$, the shockwave completely evaporates, $M(r=0)=0$ and we obtain an ordinary asymptotically flat spacetime, without horizon and singularity. In the second case $M(r)$ is almost constant for $r > 2M_{\infty}$. Evaporation starts just before the event horizon. Even though $f(r) = 1-2M(r)/r$, for $r < 2M_{\infty}$, one can check that it does not have a zero, so there is no horizon. For coordinate radii greater than the classical event horizon radius, $2M_{\infty}$, the shockwave resembles compact horizonless objects analyzed in \cite{baccetti2018role}.

The above results do not contradict the observational evidence for compact massive objects whose radii are comparable to the corresponding black hole event horizons \cite{Doeleman2008EventhorizonscaleSI}. Effects of black hole evaporation can be detected only at very late times, significantly larger than the age of the universe. $1 - M(r)/M_{\infty} \equiv \alpha$ is of order $\cO(1)$ only for $r\lesssim 2M + \delta$, with $\delta\sim \exp(-\text{const.}\times\alpha M^3 /C \hbar^2)$, which is observed far away from the black hole at time $t_{o} \sim v + 2M\log(\delta)\sim \alpha M^3/ C \hbar^2$. For one-dimensional system generic density of states is inversely proportional to $\hbar$, so $C\sim \hbar^{-1}$ and $t_{o}\sim M^3/\hbar$ is of the same order of magnitude as the standard estimate for the black hole evaporation time \cite{Hawking74, Hawking75, Lopresto:2003zz} (and much larger than the age of the universe).  

\section{Conclusions}
In the first section, we have extended the derivation of Hawking radiation by Ford \cite{ford1997quantumfieldtheorycurved} beyond the near-horizon approximation. From the calculation, one can clearly see that positive and negative frequency modes mix only along the shockwave hypersurface. In this sense, Hawking quanta are produced on the shockwave, not emitted from the black hole. We have demonstrated that the thermal spectrum of Hawking radiation can be recovered in the near-horizon limit, while our calculations further away from the horizon reveal corrections due to early-time quanta creation. 

An obvious further improvement of the method would be to go beyond the WKB approximation. That is, to study how greybody factors influence Hawking quanta created further away from the horizon. It would be also interesting to check how our modification of Bogoliubov coefficients influences the model of backreaction of \cite{Kawai:2013mda}.

Next, in sections 3 and 4 we extended the results to Reissner-Nordstr\"om and Kerr black holes with formulas also valid for superradiant modes. In particular, our analysis of extremal Reissner-Nordstrom black holes shows that while the near-horizon region does not produce a thermal flux of radiation, a finite number of Hawking quanta can still be generated further from the horizon. This confirms the results of \cite{good2020extremal}. 

Curiously, the main characteristics of our results have a simple interpretation in terms of Jacobson's toy model of Hawking radiation as the Unruh effect perceived by a distant observer \cite{Jacobson:1995ak}. In section 6 we extended this model to include the backreaction effects. A natural continuation of this work would be an extension to asymptotically dS and AdS black hole spacetimes. All in all, our simple model indicates that black holes may not form as is conventionally understood.

\section*{Acknowledgments} 
We thank Wojciech Kami\'nski and Krzysztof Meissner for helpful conversations.

\appendix
\section*{Appendix: Hawking radiation of a non-extremal RNBH}\label{Appendix_RN}
Let us now consider the formation of a non-extremal RNBH from a null, electrically charged null spherical shell. The corresponding metric is:
\be
\rd s^2 = \begin{cases}
      -\rd v^2 + 2\rd v\rd r + r^2\rd\Om^2, & \text{for}\ v<0 \\
      -f(r)\rd v^2 +2\rd v\rd r + r^2\rd\Om^2, & \text{for}\ v > 0,
    \end{cases}
\ee
where $f(r)=\frac{1}{r^2}(r-r_+)(r-r_-)$. It has two horizons for $v > 0$, at $r=r_\pm$, with surface gravities $\kappa_\pm =\frac{1}{2}f'(r_\pm)$, Tortoise coordinate:
\begin{align}
r_{\star} = \int\frac{\rd r}{f(r)} = r + \frac{1}{2\kappa_+}\log\left(\frac{r}{r_+} - 1\right) + \frac{1}{2\kappa_-}\log\left(\frac{r}{r_-}-1\right).
\end{align}
Modes
\begin{align} 
h_{\om} &= \frac{\Theta(r-r_+)}{\sqrt{4\pi\om}}\frac{Y_{lm}(\theta,\vp)}{r}e^{-i\om v + 2i\om r_\star}, 
\\
 p_{\om} &= \frac{1}{\sqrt{4\pi\om}}\frac{Y_{lm}(\theta,\vp)}{r}e^{-i\om v},
\end{align}
have positive-frequency waves near $\scri^+$ and $\scri^-$, respectively. Bogoliubov coefficients for $\om>0$ are determined from the decomposition of $h_{\om}$ in $p_{\om}$ along the shockwave $v=0$:
\begin{align}
h_\om\big|_{v=0} =\int_0^\infty \rd \om' \Big(\alpha_{\om\om'} p_{1\omega'}\big|_{v=0}+\beta_{\om\om'}p^*_{1\omega'}\big|_{v=0}\Big).\end{align}
We calculate $\beta_{\om\om'}$ coefficients noting that the above decomposition is equivalent to the Fourier transform of $r\cdot h_{\om}|_{v=0}$. Then:
\begin{align}
\nn\b_{\om\om'} = \frac{i}{2\pi}\left(\frac{\om'}{\om}\right)^{1/2} \frac{e^{i(r_+ + r_-)(\om + \om')}}{\om + \om'}\left(\frac{i(r_+ r_-)}{2r_+^2 (\om + \om')}\right)^{\frac{i\om}{2}\left(\frac{1}{\k_+} + \frac{1}{\k_-}\right)}\Gamma(1 + i\om/\k_+)\times&\\
\times W_{\frac{i\om}{2}\left(\frac{1}{\k_-} - \frac{1}{\k_+}\right), \frac{1}{2} + \frac{i\om}{2}\left(\frac{1}{\k_+} + \frac{1}{\k_-}\right)}\left(-2i(r_+ - r_-)(\om + \om')\right)&,
\end{align}
where $W_{k,m}(z)$ is the Whittaker function.
To obtain a finite expectation value of the particle number at $\scri^+$, as before, we introduce the UV cutoff:
\begin{align}
\nn\mean{N_{\om}^{\scri^+}}=&
\frac{e^{-\frac{\pi\om}{2}\left(\frac{1}{\kappa_+} + \frac{1}{\kappa_-}\right)}}{4\pi^2\om} \left|\Gamma\left(1 + \frac{i\om}{\kappa_+}\right)\right|^2\times
\\
\times&\int_0^{\Lambda}\rd\om'\frac{\om'}{(\om + \om')^2}\left|W_{\frac{i\om}{2}\left(\frac{1}{\kappa_-} - \frac{1}{\kappa_+}\right), -\frac{1}{2}- \frac{i\om}{2}\left(\frac{1}{\kappa_+} + \frac{1}{\kappa_-}\right)}\Big(-2i(r_+ - r_-)(\om + \om')\Big)\right|^2.
\end{align}
To estimate this integral we use asymptotics of Whittaker functions:
\begin{align} W_{k,m}(z)\overset{|z|\rightarrow\infty}{\sim}e^{-z/2} z^k\big(1 + \cO(1/|z|)\big),\end{align}
so that

\begin{align} \left|W_{\frac{i\om}{2}\left(\frac{1}{\kappa_-} - \frac{1}{\kappa_+}\right), -\frac{1}{2}- \frac{i\om}{2}\left(\frac{1}{\kappa_+} + \frac{1}{\kappa_-}\right)}\Big(-2i(r_+ - r_-)(\om + \om')\Big)\right|\overset{\om'\rightarrow\infty}{\sim} e^{\frac{\pi\om}{4}\left(\frac{1}{\kappa_-} -\frac{1}{\kappa_+}\right)}\left(1 + \cO(1/\om')\right).\end{align}
In this way we obtain \eqref{ReissnerNumber}:
\be
\mean{N_{\om}^{\scri^+}}= \frac{1}{2\pi\kappa_+}\frac{e^{-\frac{\pi\om}{\kappa_+}}}{e^{\frac{\pi\om}{\kappa_+}}-e^{-\frac{\pi\om}{\kappa_+}}} \int_0^{\Lambda}\rd\om'\frac{\om'}{(\om + \om')^2}
= \frac{1}{2\pi\kappa_+}\frac{1}{e^{\frac{2\pi\om}{\kappa_+}}-1} \left(\ln\Big(\frac{\Lambda}{\om}\Big)+ \cO(\Lambda^0)\right).
\ee

\bibliographystyle{utphys}
\bibliography{aipsamp}

\providecommand{\noopsort}[1]{}\providecommand{\singleletter}[1]{#1}%
\providecommand{\href}[2]{#2}\begingroup\raggedright\begin{thebibliography}{10}

\bibitem{Hawking74}
S.~W. Hawking, ``{Black hole explosions},'' \href{http://dx.doi.org/10.1038/248030a0}{{\em Nature} {\bfseries 248} (1974) 30--31}.

\bibitem{Hawking75}
S.~W. Hawking, ``{Particle Creation by Black Holes},'' \href{http://dx.doi.org/10.1007/BF02345020}{{\em Commun. Math. Phys.} {\bfseries 43} (1975) 199--220}.

\bibitem{unruh1995sonic}
W.~G. Unruh, ``Sonic analogue of black holes and the effects of high frequencies on black hole evaporation,'' {\em Phys. Rev. D} {\bfseries 51} no.~6, (1995) 2827.

\bibitem{corley1996hawking}
S.~Corley and T.~Jacobson, ``Hawking spectrum and high frequency dispersion,'' {\em Phys. Rev. D} {\bfseries 54} no.~2, (1996) 1568.

\bibitem{polchinski1995string}
J.~Polchinski, ``String theory and black hole complementarity,'' 1995.
\newblock \url{https://arxiv.org/abs/hep-th/9507094}.

\bibitem{susskind1992hawking}
L.~Susskind and L.~Thorlacius, ``Hawking radiation and back-reaction,'' {\em Nucl. Phys. B} {\bfseries 382} no.~1, (1992) 123--147.

\bibitem{Kawai:2013mda}
H.~Kawai, Y.~Matsuo, and Y.~Yokokura, ``{A Self-consistent Model of the Black Hole Evaporation},'' \href{http://dx.doi.org/10.1142/S0217751X13500504}{{\em Int. J. Mod. Phys. A} {\bfseries 28} (2013) 1350050}, \href{http://arxiv.org/abs/1302.4733}{{\ttfamily arXiv:1302.4733 [hep-th]}}.

\bibitem{hartle1976path}
J.~B. Hartle and S.~W. Hawking, ``Path-integral derivation of black-hole radiance,'' {\em Phys. Rev. D} {\bfseries 13} no.~8, (1976) 2188.

\bibitem{davies1977radiation}
P.~C.~W. Davies and S.~A. Fulling, ``{Radiation from Moving Mirrors and from Black Holes},'' \href{http://dx.doi.org/10.1098/rspa.1977.0130}{{\em Proc. Roy. Soc. Lond. A} {\bfseries 356} (1977) 237--257}.

\bibitem{robinson2005relationship}
S.~P. Robinson and F.~Wilczek, ``Relationship between hawking radiation and gravitational anomalies,'' {\em Phys. Rev. Lett.} {\bfseries 95} no.~1, (2005) 011303.

\bibitem{parikh2000hawking}
M.~K. Parikh and F.~Wilczek, ``Hawking radiation as tunneling,'' {\em Phys. Rev. Lett.} {\bfseries 85} no.~24, (2000) 5042.

\bibitem{shankaranarayanan2001method}
S.~Shankaranarayanan, K.~Srinivasan, and T.~Padmanabhan, ``Method of complex paths and general covariance of hawking radiation,'' {\em Mod. Phys. Lett. A} {\bfseries 16} no.~09, (2001) 571--578.

\bibitem{chen2019hawking}
P.~Chen, M.~Sasaki, and D.-H. Yeom, ``{Hawking radiation as instantons},'' \href{http://dx.doi.org/10.1140/epjc/s10052-019-7138-0}{{\em Eur. Phys. J. C} {\bfseries 79} no.~7, (2019) 627}, \href{http://arxiv.org/abs/1806.03766}{{\ttfamily arXiv:1806.03766 [hep-th]}}.

\bibitem{vanzo2011tunnelling}
L.~Vanzo, G.~Acquaviva, and R.~Di~Criscienzo, ``Tunnelling methods and hawking's radiation: achievements and prospects,'' {\em Classical Quantum Gravity} {\bfseries 28} no.~18, (2011) 183001.

\bibitem{helfer2019hawking}
A.~D. Helfer, ``Hawking radiation, quantum fields, and tunneling,'' {\em Phys. Rev. D} {\bfseries 100} no.~2, (2019) 025005.

\bibitem{ford1997quantumfieldtheorycurved}
L.~H. Ford, ``Quantum field theory in curved spacetime,'' 1997.
\newblock \url{https://arxiv.org/abs/gr-qc/9707062}.

\bibitem{Wald_book}
R.~M. Wald, {\em Quantum field theory in curved spacetime and black hole thermodynamics}.
\newblock University of Chicago Press, 1994.

\bibitem{birrell_davies_1982}
N.~D. Birrell and P.~C.~W. Davies, \href{http://dx.doi.org/10.1017/CBO9780511622632}{{\em Quantum Fields in Curved Space}}.
\newblock Cambridge Monographs on Mathematical Physics. Cambridge University Press, 1982.

\bibitem{parker2009quantum}
L.~Parker and D.~Toms, {\em Quantum field theory in curved spacetime: quantized fields and gravity}.
\newblock Cambridge University Press, 2009.

\bibitem{unruh1976notes}
W.~G. Unruh, ``Notes on black-hole evaporation,'' {\em Phys. Rev. D} {\bfseries 14} no.~4, (1976) 870.

\bibitem{mersini2014backreaction}
L.~Mersini-Houghton, ``Backreaction of hawking radiation on a gravitationally collapsing star i: Black holes?'' {\em Phys. Lett. B} {\bfseries 738} (2014) 61--67.

\bibitem{baccetti2018role}
V.~Baccetti, R.~B. Mann, and D.~R. Terno, ``Role of evaporation in gravitational collapse,'' {\em Classical Quantum Gravity} {\bfseries 35} no.~18, (2018) 185005.

\bibitem{chen2018pre}
P.~Chen, W.~G. Unruh, C.-H. Wu, and D.-H. Yeom, ``Pre-hawking radiation cannot prevent the formation of apparent horizon,'' {\em Phys. Rev. D} {\bfseries 97} no.~6, (2018) 064045.

\bibitem{unruh2018prehawking}
W.~G. Unruh, ``Prehawking radiation,'' 2018.
\newblock \url{https://arxiv.org/abs/1802.09107}.

\bibitem{barcelo2011minimal}
C.~Barcelo, S.~Liberati, S.~Sonego, and M.~Visser, ``Minimal conditions for the existence of a hawking-like flux,'' {\em Phys. Rev. D} {\bfseries 83} no.~4, (2011) 041501.

\bibitem{Shale1962LINEARSO}
D.~Shale, ``Linear symmetries of free boson fields,'' {\em Trans.Amer.Math.Soc} {\bfseries 103} no.~1, (1962) 149--167. \url{http://www.jstor.org/stable/1993745}.

\bibitem{hooft1985quantum}
G.~'t~Hooft, ``{On the Quantum Structure of a Black Hole},'' \href{http://dx.doi.org/10.1016/0550-3213(85)90418-3}{{\em Nucl. Phys. B} {\bfseries 256} (1985) 727--745}.

\bibitem{FredenhagenHaag}
K.~Fredenhagen and R.~Haag, ``{On the Derivation of Hawking Radiation Associated With the Formation of a Black Hole},'' \href{http://dx.doi.org/10.1007/BF02096757}{{\em Commun. Math. Phys.} {\bfseries 127} (1990) 273}.

\bibitem{ryzhikSpecial}
I.~S. Gradshteyn and I.~M. Ryzhik, {\em Table of integrals, series, and products}.
\newblock Academic Press, 2014.

\bibitem{kehle2024extremal}
C.~Kehle and R.~Unger, ``Extremal black hole formation as a critical phenomenon,'' 2024.
\newblock \url{https://arxiv.org/abs/2402.10190}.

\bibitem{Angheben:2005}
M.~Angheben, M.~Nadalini, L.~Vanzo, and S.~Zerbini, ``{Hawking radiation as tunneling for extremal and rotating black holes},'' \href{http://dx.doi.org/10.1088/1126-6708/2005/05/014}{{\em J. High Energy Phys.} {\bfseries 05} (2005) 014}, \href{http://arxiv.org/abs/hep-th/0503081}{{\ttfamily arXiv:hep-th/0503081}}.

\bibitem{good2020extremal}
M.~R. Good, ``Extremal hawking radiation,'' {\em Phys. Rev. D} {\bfseries 101} no.~10, (2020) 104050.

\bibitem{NewmanJanis1965}
E.~T. Newman and A.~I. Janis, ``{Note on the Kerr spinning particle metric},'' \href{http://dx.doi.org/10.1063/1.1704350}{{\em J. Math. Phys.} {\bfseries 6} (1965) 915--917}.

\bibitem{Dahal:2020KerrVaidya}
P.~K. Dahal and D.~R. Terno, ``{Kerr-Vaidya black holes},'' \href{http://dx.doi.org/10.1103/PhysRevD.102.124032}{{\em Phys. Rev. D} {\bfseries 102} (2020) 124032}, \href{http://arxiv.org/abs/2008.13370}{{\ttfamily arXiv:2008.13370 [gr-qc]}}.

\bibitem{Bicak:1993ThinDisk}
J.~Bicak and T.~Ledvinka, ``{Relativistic disks as sources of the Kerr metric},'' \href{http://dx.doi.org/10.1103/PhysRevLett.71.1669}{{\em Phys. Rev. Lett.} {\bfseries 71} (1993) 1669--1672}.

\bibitem{carter1979generalized}
B.~Carter and R.~G. McLenaghan, ``Generalized total angular momentum operator for the dirac equation in curved space-time,'' {\em Phys. Rev. D} {\bfseries 19} no.~4, (1979) 1093.

\bibitem{teukolsky1973perturbations}
S.~A. Teukolsky, ``Perturbations of a rotating black hole. i. fundamental equations for gravitational, electromagnetic, and neutrino-field perturbations,'' {\em Astrophys. J.} {\bfseries 185} (1973) 635--648.

\bibitem{press1973perturbations}
W.~H. Press and S.~A. Teukolsky, ``Perturbations of a rotating black hole. ii. dynamical stability of the kerr metric,'' {\em Astrophys. J.} {\bfseries 185} (1973) 649--674.

\bibitem{teukolsky1974perturbations}
S.~A. Teukolsky and W.~H. Press, ``Perturbations of a rotating black hole. iii-interaction of the hole with gravitational and electromagnetic radiation,'' {\em Astrophys. J.} {\bfseries 193} (1974) 443--461.

\bibitem{AbramowitzStegun}
M.~Abramowitz, I.~A. Stegun, and D.~Miller, ``Handbook of mathematical functions with formulas, graphs and mathematical tables (national bureau of standards applied mathematics series no. 55),'' {\em J. Appl. Mech.} {\bfseries 32} (1965) 239--239. \url{https://api.semanticscholar.org/CorpusID:121782574}.

\bibitem{MaldacenaStrominger1997}
J.~M. Maldacena and A.~Strominger, ``{Universal low-energy dynamics for rotating black holes},'' \href{http://dx.doi.org/10.1103/PhysRevD.56.4975}{{\em Phys. Rev. D} {\bfseries 56} (1997) 4975--4983}, \href{http://arxiv.org/abs/hep-th/9702015}{{\ttfamily arXiv:hep-th/9702015}}.

\bibitem{Jacobson:1995ak}
T.~A. Jacobson, ``Introduction to black hole microscopy,'' 1995.
\newblock \url{https://arxiv.org/abs/hep-th/9510026}.

\bibitem{Barcelo:2010PreHawkingMinimalConditions}
C.~Barcelo, S.~Liberati, S.~Sonego, and M.~Visser, ``{Minimal conditions for the existence of a Hawking-like flux},'' \href{http://dx.doi.org/10.1103/PhysRevD.83.041501}{{\em Phys. Rev. D} {\bfseries 83} (2011) 041501}, \href{http://arxiv.org/abs/1011.5593}{{\ttfamily arXiv:1011.5593 [gr-qc]}}.

\bibitem{mersini2014back}
L.~Mersini-Houghton, ``Back-reaction of the hawking radiation flux on a gravitationally collapsing star ii,'' 2015.
\newblock \url{https://arxiv.org/abs/1409.1837}.

\bibitem{Doeleman2008EventhorizonscaleSI}
S.~S. Doeleman, J.~Weintroub, A.~E.~E. Rogers, R.~L. Plambeck, R.~W. Freund, R.~P.~J. Tilanus, P.~Friberg, L.~M. Ziurys, J.~M. Moran, B.~E. Corey, K.~Young, D.~L. Smythe, M.~Titus, D.~P. Marrone, R.~J. Cappallo, D.~C.-J. Bock, G.~C. Bower, R.~A. Chamberlin, G.~R. Davis, T.~P. Krichbaum, J.~W. Lamb, H.~L. Maness, A.~E. Niell, A.~L. Roy, P.~A. Strittmatter, D.~Werthimer, A.~R. Whitney, and D.~P. Woody, ``Event-horizon-scale structure in the supermassive black hole candidate at the galactic centre,'' {\em Nature} {\bfseries 455} (2008) 78--80. \url{https://api.semanticscholar.org/CorpusID:4424735}.

\bibitem{Lopresto:2003zz}
M.~C. Lopresto, ``{Some Simple Black Hole Thermodynamics},'' \href{http://dx.doi.org/10.1119/1.1571268}{{\em Phys. Teacher} {\bfseries 41} (2003) 299--301}.

\end{thebibliography}\endgroup
\end{document}